\documentclass[twocolumn,tighten]{aastex63}
\accepted{}
\received{}
\submitjournal{AAS journals}
\usepackage{graphicx}
\usepackage{natbib}
\usepackage{latexsym}
\usepackage{amssymb}
\usepackage{longtable}
\usepackage{amsmath}
\usepackage{url}
\def\msun{\ifmmode {\rm\,M_\odot}\else ${\rm\,M_\odot}$\fi}
\def\Msun{\ifmmode {\rm\,\it{M_\odot}}\else ${\rm\,M_\odot}$\fi}
\def\lsun{\ifmmode {\rm\,L_\odot}\else ${\rm\,L_\odot}$\fi}
\def\Lsun{\ifmmode {\rm\,\it{L_\odot}}\else ${\rm\,L_\odot}$\fi}
\def\rsun{\ifmmode {\rm\,R_\odot}\else ${\rm\,R_\odot}$\fi}
\def\Rsun{\ifmmode {\rm\,\it{R_\odot}}\else ${\rm\,R_\odot}$\fi}
\def\Tsun{\ifmmode {\rm\,T_\odot}\else ${\rm\,T_\odot}$\fi}
\def\arcsec{\ifmmode {^{\prime\prime}}\else $^{\prime\prime}$\fi}
\def\asec{\ifmmode {^{\prime\prime}}\else $^{\prime\prime}$\fi}
\def\arcmin{\ifmmode {^{\prime}}\else $^{\prime}$\fi}
\def\amin{\ifmmode {^{\prime}}\else $^{\prime}$\fi}
\def\simlt{\mathrel{\spose{\lower 3pt\hbox{$\mathchar"218$}}
     \raise 2.0pt\hbox{$\mathchar"13C$}}}
\def\simgt{\mathrel{\spose{\lower 3pt\hbox{$\mathchar"218$}}
\     \raise 2.0pt\hbox{$\mathchar"13E$}}}

\def\escape~{\textit{ESCAPE}}

\begin{document}

\author[0000-0002-1002-3674]{Kevin France}
\affiliation{Laboratory for Atmospheric and Space Physics, University of Colorado Boulder, Boulder, CO 80309}
\affiliation{Department of Astrophysical and Planetary Sciences, University of Colorado Boulder, Boulder, CO 80309}
\affiliation{Center for Astrophysics and Space Astronomy, University of Colorado Boulder, Boulder, CO 80309}

\author[0000-0002-7119-2543]{Girish Duvvuri}
\affiliation{Department of Physics and Astronomy, Vanderbilt University, Nashville, TN 37235, USA}

\author[0000-0001-8499-2892]{Cynthia S. Froning}
\affiliation{Southwest Research Institute, San Antonio, TX 78238}

\author[0000-0003-2631-3905]{Alexander Brown}
\affiliation{Center for Astrophysics and Space Astronomy, University of Colorado Boulder, Boulder, CO 80309}

\author[0000-0002-5094-2245]{P. Christian Schneider}
\affiliation{Christian-Albrechts University, Leibnizstra\ss{}e 15, 
24118 Kiel, Germany}

\author[0000-0002-4489-0135]{J.\ Sebastian Pineda}
\affiliation{Laboratory for Atmospheric and Space Physics, University of Colorado Boulder, Boulder, CO 80309}

\author[0000-0001-9667-9449]{David Wilson}
\affiliation{Laboratory for Atmospheric and Space Physics, University of Colorado Boulder, Boulder, CO 80309}

\author[0000-0002-1176-3391]{Allison Youngblood}
\affiliation{Exoplanets and Stellar Astrophysics Laboratory, NASA Goddard Space Flight Center, Greenbelt, MD 20771}

\author[0000-0003-4452-0588]{Vladimir S. Airapetian}
\affiliation{NASA GSFC/SEEC, Greenbelt, MD, USA}
\affiliation{Department of Physics, American University, Washington, DC, USA}

\author[0000-0002-1297-9485]{Kosuke Namekata}
\affil{Heliophysics Science Division, NASA Goddard Space Flight Center, 8800 Greenbelt Road, Greenbelt, MD 20771, USA}
\affiliation{The Catholic University of America, 620 Michigan Avenue, N.E. Washington, DC 20064, USA}
\affiliation{The Hakubi Center for Advanced Research, Kyoto University, Yoshida-Honmachi, Sakyo-ku, Kyoto 606-8501, Japan}
\affiliation{Department of Physics, Kyoto University, Kitashirakawa-Oiwake-cho, Sakyo-ku, Kyoto, 606-8502, Japan}

\author[0000-0002-0412-0849]{Yuta Notsu}
\affiliation{Laboratory for Atmospheric and Space Physics, University of Colorado Boulder, Boulder, CO 80309}
\affiliation{Department of Astrophysical and Planetary Sciences, University of Colorado Boulder, Boulder, CO 80309}
\affil{National Solar Observatory, Boulder, CO 80309, USA}

\author[0009-0006-0318-3385]{Tristen Sextro}
\affiliation{Department of Astronomy and Astrophysics, Pennsylvania State University, University Park, PA 16802, USA}

\correspondingauthor{Kevin France}
\email{kevin.france@colorado.edu}

\title{A Semi-Empirical Estimate of Solar EUV Evolution from 10 Myr to 10 Gyr}

\begin{abstract} 

The extreme-ultraviolet (EUV; 100~--~911~\AA) spectra of F, G, K, and M stars provide diagnostics of the stellar chromosphere through the corona, with line and continuum formation temperatures spanning roughly 10$^{4}$ - 10$^{7}$ K.   The EUV stellar spectrum in turn drives atmospheric photochemistry and numerous escape processes on orbiting planets, and is often combined with the stellar X-ray flux to make up the XUV irradiance.  However, very few direct EUV spectra of other stars exist in the archive and as a result, X-ray scaling relations are often assumed for the XUV evolution of cool stars.  In this work, we present a new study of the EUV history of solar-type stars, using scaling relations based on transition region/coronal far-ultraviolet emission lines and differential emission measure (DEM)-based synthetic spectra to provide a semi-empirical estimate of the EUV evolution of the Sun over the $\approx$~10 Myr~to~10 Gyr age range for the first time.  We utilize new and archival {\it Hubble Space Telescope} observations of solar analogs (T$_{\odot}$~$\pm$~150 K for stars older than 100 Myr) and ``Young Suns" (age $<$ 100 Myr) that will evolve into main sequence early G-type stars to predict the 90~--~360~\AA\ EUV flux from a sample of 23 stars.  We find that the EUV activity evolution for solar-type stars follows a two-component behavior: a saturated L(EUV)/L$_{bol}$ plateau (at a level of about 10$^{-4}$) followed by a power law decay ($\alpha$ $\approx$~$-$1.1) after ages of $\approx$~50~--~100 Myr.   Consequently, the EUV flux incident at 1 AU around solar analogs varies over the lifetime of the Sun, ranging from 100~$\times$ the present day UV irradiance at 10 Myr to 0.3~$\times$ the present-day level at 10 Gyr.   We find that the EUV luminosity is approximately the same as the soft X-ray luminosity up to approximately 1 Gyr, after which the EUV luminosity of the stars dominate.
In comparison to Sun-like stars, the EUV saturation level of early/mid M dwarfs is several times higher and lasts $\sim$10~--~20 times longer.

\end{abstract}



\section{INTRODUCTION}
\label{sec:intro} 

The stability of Earth-like atmospheres depends strongly on the high-energy irradiance from the parent star~\citep{johnstone18,airapetian20,gronoff2020}.  Elevated extreme-ultraviolet (EUV; 100~--~911~\AA) fluxes of 5 – 10 times the present Earth-Sun level, representative of earlier phases in the Sun's history, are predicted to drive rapid mass loss from terrestrial atmospheres~\citep{tu15,nakayama22}. Additionally, free electrons produced by stellar EUV photons can attain altitudes much greater than ions, producing an ambipolar electric field that leads to a non-thermal ionospheric outflow (O$^{+}$ and N$^{+}$ winds; \citealt{kulikov06,dong17,airapetian17}), the dominant source of atmospheric loss from Earth today~\citep{seki01}. The stellar EUV environment is therefore an essential input to atmospheric escape estimates for both terrestrial (e.g., \citealt{nakayama22,zhang24}) and gaseous (e.g., \citealt{koskinen22,zhang22}) planets.  This energy input likely played a significant role in the evolution of planets in our own solar system (e.g., Mars; \citealt{jakosky18}) and fundamentally shifts our interpretation of and predictions for which exoplanets likely host long-lived habitable atmospheres.  These predictions are consequently limited by the uncertainties associated with the stellar EUV environment.

The combination of high EUV luminosity and frequent impacts may have driven the loss of most volatiles from primary terrestrial atmospheres in the solar system’s first few hundred Myr~\citep{zahnle10,wordsworth13}, although the scenario for primary atmosphere retention/loss is less clear for rocky planets orbiting M dwarfs (see, e.g., \citealt{totton24} and references therein).   Potentially habitable M dwarf exoplanets are particularly prone to EUV-driven atmospheric escape owing to their close-in orbits \citep{youngblood16,airapetian17,france20}.  Additionally, the EUV luminosity of M dwarfs is thought to be enhanced by another factor of $\sim$~10 relative to their main sequence level during their prolonged pre-main-sequence evolution and slow rotational spin-down, which can last up to several Gyrs longer than that of Sun-like stars~\citep{pineda21,loyd21,pass24}, well into the era in which life emerged on Earth.    

This is being investigated through observations of small (R$_{P}$ $<$4 R$_{Earth}$), hotter ($T_{eff}$~$>$~500 K), exoplanets orbiting M dwarfs interior to the habitable zone with $JWST$ (e.g., \citealt{cadieux24,august25}).  Numerous studies suggest that at least some rocky planets on close-in orbits around M dwarfs have suffered complete atmospheric removal ~\citep{kreidberg19,crossfield22,greene23,zieba23,wachiraphan25} or are consistent with high mean molecular weight atmospheres, cloudy atmospheres, or no atmospheres ~\citep{lustig23,moran23,may23,lim23,mansfield24}.  Larger samples of cooler, rocky planets are being surveyed by $JWST$ today, and these results will shape our understanding of atmospheric survival around M dwarfs \citep{redfield24} into the 2030s.

As the community looks ahead to the discovery and characterization of Earth-Sun analogs with the Habitable Worlds Observatory (HWO; \citealt{arney25}), a better understanding of the EUV history of F, G, and K type stars becomes important for understanding the potential for these worlds to maintain habitable conditions.  Are rocky planets orbiting K and late G-type stars able to support long-lived atmospheres or do they follow the apparent fate of most M dwarf planets?  We require a better quantitative understanding of 
the cumulative high-energy inputs into exoplanetary atmospheres~\citep{claire12,luger15} to empirically determine the location of the ``cosmic shoreline''~\citep{zahnle17}.  Questions regarding whether K stars behave more like M stars or solar-type stars (see, e.g., \citealt{yowell22}) and which solar-type stars share a similar cumulative high-energy input (e.g., \citealt{johnstone21}) will help the community to identify the star-planet systems most conducive to long-term atmospheric stability and likely habitability.  Towards this end, we present a new analysis of the EUV history of solar analogs, from the earliest ages following the dispersal of the protoplanetary disk to the oldest solar-type stars in our local Galactic neighborhood.

\subsection{EUV, XUV, and the Cosmic Shoreline}

The concept of the ``cosmic shoreline'' as laid out by~\citet{zahnle17} compares the ability of a planet to gravitationally retain an atmosphere (usually expressed as the planetary escape velocity) with the cumulative X-ray + EUV ( = XUV) energy that the host star deposits into a planet's atmosphere over its lifetime.  While X-rays contribute to the ionization and heating of the middle atmosphere and are often readily observable, EUV photons dominate the energy input, photoelectron production, and temperature structure of the upper atmospheres of terrestrial planets orbiting both active and inactive stars~\citep{solomon05,youngblood25,vanLooveren24}.    EUV photons are absorbed in the highest (lowest density) layers of the atmosphere where radiative losses are minor and the heating efficiency is highest. While the integrated X-ray luminosity is thought to be larger than the integrated EUV luminosity for young stars (althopugh see Section 3.4), the total number of  EUV photons is larger (3-90$\times$) on stars of all ages~\citep{woods09,fontenla16,garciasage17,king21}. 

Despite their importance, EUV spectra of exoplanet host stars are scarce. The only previous EUV astronomy mission, $EUVE$~\citep{bowyer91}, obtained (typically low-S/N) spectra of approximately a dozen cool main sequence stars, including 5 solar analogs~\citep{craig97, ribas05} and 5 M dwarfs.  

In practice, the \citet{zahnle17} ``cosmic shoreline'' calculation computes XUV evolutionary histories from X-ray measurements~\citep{lammer09, penz08a,penz08b}, described by a saturated early phase followed by a power-law decline.  This behavior is well-established for the X-ray history of solar type stars, but the situation for the EUV regime is much less empirically grounded: the saturated phase has never been empirically observed and the slope in the declining phase is based on measurements of 4 stars, including the Sun.  Many more X-ray and far-ultraviolet observations of solar-type stars (in particular stars younger than 100 Myr) are available today compared to the pioneering `Sun in Time' study of~\citet{ribas05}.  The goal of this work is to present a new analysis of the EUV history of solar-type stars by combining empirically-derived FUV-to-EUV scaling relations and differential emission measure (DEM)-based EUV calculations (Section 2.4).

This paper is laid out as follows:  Section 2 describes the sample of solar-like stars, the methodologies we adopt for EUV emission calculations, and the new and archival data used in this study.   Section 3 presents the overall empirical results of this work: the time evolution of the absolute and fractional (normalized by the bolometric flux) EUV luminosities of our sample, the evolution of the EUV flux in the habitable zone, and a comparison with the EUV evolution of M dwarfs.    We conclude in Section 4 and discuss the remaining (substantial) uncertainties associated with the paucity of direct EUV observations of cool stars.   The Appendix explores a likely FUV superflare caught on a $\approx$ 7 Myr star.

\begin{figure}[htbp]
   \centering
   \includegraphics[scale=.6,clip,trim=0mm 0mm 0mm 0mm,angle=0]{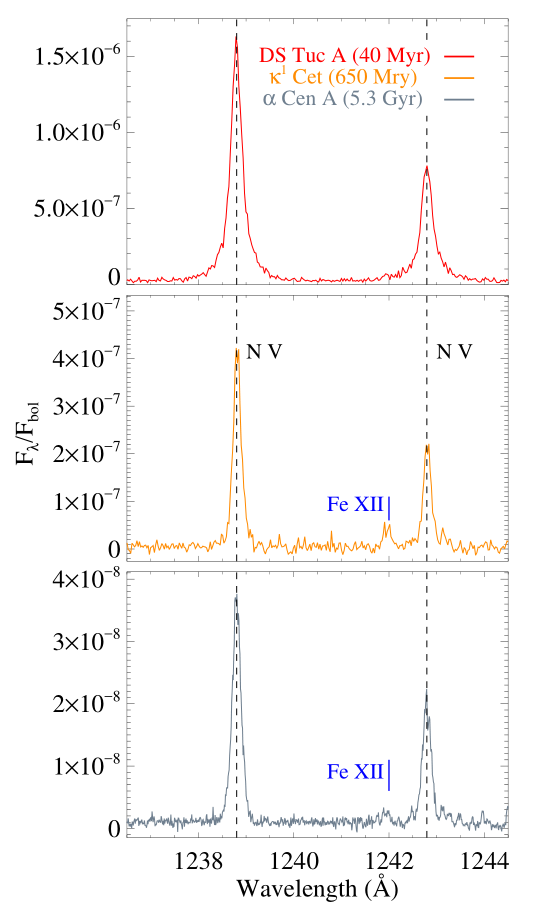}
   \figcaption{Representative \ion{N}{5} data from $HST$-COS (top) and STIS (middle and bottom) roughly spanning the age range presented in this work.  From top to bottom, the histogram shows the bolometric flux-normalized \ion{N}{5} spectra (tracing transition region formation temperatures $\approx$~1~--~2~$\times$~10$^{5}$ K) of DS Tuc A ($\sim$~40 Myr; red), $\kappa^{1}$~Cet ($\sim$~650 Myr; orange), and $\alpha$~Cen A ($\sim$~5.3 Gyr; gray).  The spectra have been velocity-shifted to place \ion{N}{5} at its rest wavelength (noted with dashed vertical lines) for comparison, and weak coronal \ion{Fe}{12} is noted with a blue label in $\kappa^{1}$~Cet and $\alpha$~Cen A.}
\end{figure}

\begin{figure}[htbp]
   \centering
   \hspace{-0.3in}
   \includegraphics[scale=.45,clip,trim=0mm 0mm 0mm 0mm,angle=0]{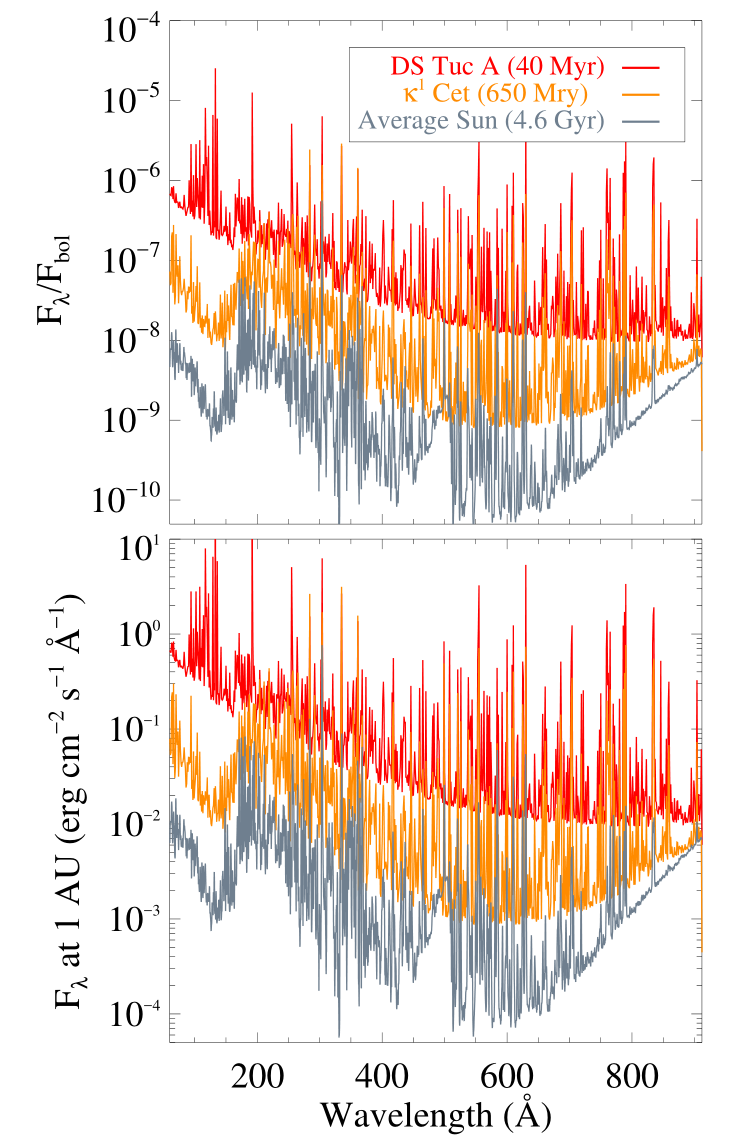}
   \figcaption{Representative DEM-based synthetic EUV spectra and direct EUV observations (shown here from 60~--~912~\AA) roughly spanning the age range presented in this work.  In the top plot, the histogram shows the bolometric flux-normalized EUV spectra of DS Tuc A ($\sim$~40 Myr; red), $\kappa^{1}$~Cet ($\sim$~650 Myr; orange), and the Sun at moderate activity levels ($\sim$~4.6 Gyr; gray).  The DS Tuc and $\kappa^{1}$~Cet synthetic spectra are developed using the formalism described in \citet{duvvuri21}, and the solar data are from the SDO-EVE instrument~\citep{woods12}.  The bottom plot shows the absolute flux of the same spectra, as received at 1 astronomical unit from the star.}
\end{figure}

\section{Solar-analog sample, EUV calculations, and Observations}

\subsection{An expanded sample of Solar-analogs with high-quality FUV observations}

We present a new sample of 23 ``solar-type" stars, spanning a wide range of ages.   Defining a set of criteria for what constitutes a solar-type star is somewhat subjective (see, e.g., \citealt{deMello14}), so we balance narrowness around a 1.0 solar masses and the Sun's 5770 K effective temperature with what data are available in the literature.   For this study, we restrict our main sequence sample to stars with temperatures within 150 K of the solar value; our sample of stars (older than 100 Myr) have an average effective temperature of 5812 K with a standard deviation of 75 K.  In practice, this incorporates most main sequence G1, G2, and G3 stars with far-UV spectra (1150~--~1450~\AA) from $HST$.   As rotation periods are not available for approximately half the stars in the sample, we adopt stellar ages from a variety of literature sources. The youngest stars in the sample are based on cluster/association properties~\citep{pecaut16,david19,newton19,rizzuto20}.  Most of the ages for stars older than~$\sim$~100 Myr are derived from age-activity relationships based on \ion{Ca}{2} $R^{'}_{HK}$~\citep{rochapinto04,mamajek08,gondoin20}, FUV or X-ray observations~\citep{penz08a,crandall20}, or age-abundance isochrones~\citep{nissen15,stanfordmoore20}.  While there is no age-metallicity correlation for nearby solar-type stars (see, e.g., \citep{gondoin20} and references therein), chromospheric age determination can be affected by metallicity~\citep{pinto98,lorenzo16}; as a subset of our sample stars do not have measured metallicity, this is a potential source of uncertainty in the evolutionary curves presented here.   

We note that for some of the well-studied stars in our sample, there are modest discrepancies in the retrieved ages.  The most noteworthy is $\kappa^{1}$ Cet, for which \ion{Ca}{2} $R^{'}_{HK}$-based ages find $\approx$~350 Myr~\citep{mamajek08} while X-ray-based ages find $\approx$~750 Myr~\citep{telleschi05}; we adopt the more canonically used 650 Myr~\citep{ribas05} for this study. Most ages in the above sources do not include uncertainties, so we did not attempt to reproduce them here, but do note that the points naturally fall on an age/high-energy luminosity relationship as expected for cool star coronal activity (Section 3.1).  
 
As young stars descend on to the main sequence, their radii contract and their temperatures increase.  Therefore, as we look back to stars younger than $\sim$~100 Myr, later spectral types represent the precursors of main sequence solar-type stars.   This assignment of `protostellar' becomes more challenging for younger and younger stars.  As only a very small number of post-mass accretion stars younger than 100 Myr have existing UV characterization, we include these to complete the sample.  We acknowledge that spectral type / reddening degeneracy and uncertainties in pre-main sequence evolutionary tracks make a firm determination of the best `Young Suns' subject to uncertainty.  However, as we will show in Section 3, all of the young stars are consistent in both absolute and fractional EUV luminosity in a `saturated' age group younger than $\sim$~100~Myr, so this choice is not expected to significantly bias the outcomes of this study.

\begin{longrotatetable}
\begin{deluxetable*}{lccccccccccc}
\tabletypesize{\scriptsize}
\tablewidth{0pt}
\tablecaption{Solar-type targets and EUV Properties}
\tablehead{
\colhead{Name} & \colhead{Age} & \colhead{Distance} & \colhead{T$_{eff}$} & \colhead{f$_{Bol}$}  & \colhead{F(NV)}  &  \colhead{$Err$(NV)} &   \colhead{L(EUV)/L$_{bol}$}  &  \colhead{$Err$(L(EUV)/L$_{bol}$)}   &  \colhead{L(EUV)} &  \colhead{$Err$(L(EUV))}   &  \colhead{Notes} \\
\colhead{} & \colhead{(Gyr)} & \colhead{(pc)} & \colhead{(K)} & \colhead{(erg cm$^{-2}$ s$^{-1}$)}  & \colhead{(erg cm$^{-2}$ s$^{-1}$)} & \colhead{(erg cm$^{-2}$ s$^{-1}$)}  &  \colhead{ }   &   \colhead{ }  & \colhead{(erg s$^{-1}$) }  &  \colhead{(erg s$^{-1}$) }  &    \colhead{ }
}
\startdata
2MASS-J16081474-1908327 & 0.006 & 139.8 & $\cdots$ & 1.81E-09 & 2.16E-15 &  0.2E-15 & 9.48E-05 &  3.50E-05 & 4.01E+29 &  1.48E+29 &  A$_{V}$=0.57, \\
 &  &  &  &  &  &  &  &  &  &  &  M$_{*}$=1.3M$_{\odot}$, (1)  \\
2MASS-J16025123-2401574 & 0.006 & 145.0 & $\cdots$ & 9.25E-10 & 1.19E-15 & 0.1E-15 & 1.02E-04 & 3.76E-05  & 2.38E+29 &  8.76E+28 &   A$_{V}$=0.2, \\
\\
 &  &  &  &  &  &  &  &  &  &  &  M$_{*}$=1.0M$_{\odot}$, (1)  \\
2MASS-J16193396-2228294 & 0.008 & 128.2 & $\cdots$ & 2.45E-09 & 4.50E-15 & 0.4E-15 & 1.46E-04 & 5.39E-05 & 7.04E+29 &    2.59E+29 &  A$_{V}$=0.47, \\
\\
 &  &  &  &  &  &  &  &  &  &  &  M$_{*}$=1.3M$_{\odot}$, (1)  \\
HIP67522-NV & 0.017 & 124.7 & 5675 & 3.59E-09 & 5.89E-15 & 0.6E-15 & 1.30E-04 & 4.81E-05 & 8.71E+29 &  3.21E+29 &   (2)  \\
HIP67522-DEM & 0.017 & 124.7 & 5675 & 3.59E-09 & $\cdots$ & $\cdots$ & 1.68E-04 & 5.03E-05 & 1.12E+30 &    3.36E+29 & (2)  \\
V1298Tau-NV & 0.023 & 108.0 & 4970 & 2.56E-09 & 3.34E-15 & 0.3E-15 & 1.04E-04 & 3.83E-05 & 3.70E+29 &    1.37E+29  & (3)  \\
V1298Tau-DEM & 0.023 & 108.0 & 4970 & 2.56E-09 & $\cdots$ & $\cdots$ & 7.16E-05 & 2.15E-05 & 2.55E+29 &    7.66E+28 & (3)  \\
DSTuc-NV & 0.04 & 44.2 & 5428 & 1.19E-08 & 1.06E-14 & 1.1E-15 & 7.10E-05 & 2.62E-05 & 1.97E+29 &    7.25E+28 & (4)  \\
DSTuc-DEM & 0.04 & 44.2 & 5428 & 1.19E-08 & $\cdots$ & $\cdots$ & 1.62E-04 & 4.85E-05 & 4.48E+29 &    1.34E+29 & (4, 5)  \\
EKDra-NV & 0.1 & 34.4 & 5870 & 2.46E-08 & 2.90E-14 & 2.9E-15 & 9.38E-05 & 3.46E-05 & 3.26E+29 &    1.20E+29 & (6)  \\
EKDra-DEM & 0.1 & 34.4 & 5870 & 2.46E-08 & $\cdots$ & $\cdots$ & 1.90E-04 & 5.71E-05 & 6.61E+29 &    1.98E+29 & (6)  \\
$\pi^{1}$UMa & 0.25 & 14.4 & 5850 & 1.48E-07 & 3.11E-14 & 3.1E-15 & 1.67E-05 & 6.91E-06 & 6.16E+28 &    2.55E+28 & (7)  \\
$\chi^{1}$Ori & 0.35 & 8.7 & 5890 & 4.30E-07 & 5.53E-14 & 5.5E-15 & 1.02E-05 & 3.77E-06 & 3.94E+28 &   1.46E+28 & (7)  \\
HD62850 & 0.4 & 32.9 & 5880 & 3.53E-08 & 9.67E-15 & 1.0E-15 & 2.18E-05 & 8.03E-06 & 9.95E+28 &    3.67E+28 & (8,9)  \\
$\kappa^{1}$Cet-NV & 0.65 & 9.3 & 5750 & 2.97E-07 & 4.00E-14 & 4.0E-15 & 1.07E-05 & 3.95E-06 & 3.27E+28 &   1.21E+28 &  (6,8,10)  \\
$\kappa^{1}$Cet-DEM & 0.65 & 9.3 & 5750 & 2.97E-07 & $\cdots$ & $\cdots$ & 2.09E-05 & 6.26E-06 & 6.38E+28 &    1.91E+28 & (6,8,10)  \\
HD150706 & 0.9 & 28.2 & 5920 & 4.18E-08 & 4.18E-15 & 0.4E-15 & 7.94E-06 & 2.93E-06 & 3.16E+28 &   1.16E+28 &  (11,9)  \\
HD172669 & 2.0 & 36.0 & 5950 & 2.48E-08 & 8.43E-16 & 0.8E-16 & 2.70E-06 & 9.96E-07 & 1.04E+28 &    3.83E+27 & (12a,b,9)  \\
HD180712 & 3.0 & 45.1 & 5880 & 1.71E-08 & 4.96E-16 & 0.5E-16 & 2.30E-06 & 8.50E-07 & 9.59E+27 &    3.54E+27 & (13,14,9)  \\
HD24293 & 3.5 & 42.3 & 5730 & 1.96E-08 & 2.55E-16 & 0.3E-16 & 1.03E-06 & 3.81E-07 & 4.34E+27 &    1.60E+27 & (15,9)  \\
HD144061 & 3.6 & 29.6 & 5680 & 3.34E-08 & 3.67E-16 & 0.4E-16 & 8.73E-07 & 3.22E-07 & 3.06E+27 &    1.12E+27 & (13,14,9)  \\
Sun(MIN) & 4.6 & 0.0 & 5772 & 1.36E+06 & 1.70E-02 & $\cdots$ & 9.93E-07 & 3.66E-07 & 3.80E+27 &    1.40E+27 & (17)  \\
Sun(MAX) & 4.6 & 0.0 & 5772 & 1.36E+06 & 2.26E-02 & $\cdots$  & 1.32E-06 & 4.87E-07 & 5.05E+27 &   1.86E+27 &  (17)  \\
HD128620 & 5.3 & 1.4 & 5800 & 2.72E-05 & 4.10E-13 & 4.1E-14 & 1.20E-06 & 4.42E-07 & 7.10E+27 &    2.62E+27 & (7)  \\
HD28471 & 7.0 & 43.7 & 5750 & 1.88E-08 & 1.88E-16 & 0.2E-16 & 7.94E-07 & 2.93E-07 & 3.41E+27 &    1.26E+27 & (16,9)  \\
HD73744 & 7.5 & 36.6 & 5850 & 2.47E-08 & 3.71E-16 & 0.4E-16 & 1.19E-06 & 4.40E-07 & 4.72E+27 &    1.74E+27 & (15,9)  \\
HD32778 & 8.2 & 22.3 & 5720 & 4.36E-08 & 2.18E-16 & 0.2E-16 & 3.97E-07 & 1.47E-07 & 1.03E+27 &    3.80E+26 & (15,9)  \\
HD51929 & 10.6 & 37.4 & 5800 & 3.02E-08 & 1.21E-16 & 0.1E-16 & 3.18E-07 & 1.17E-07 & 1.61E+27 &    5.93E+26 & (15,9)  \\
HD28701 & 11.2 & 44.2 & 5750 & 1.96E-08 & 7.84E-17 & 0.1E-16 & 3.18E-07 & 1.17E-07 & 1.46E+27 &    5.37E+26 & (15,9) 
\enddata
\tablecomments{References for the stellar parameters and archival data:
(1) \citep{pecaut16}, 
(2) \citep{rizzuto20},  
(3) \citep{david19}, 
(4) \citep{newton19}, 
(5) (dos Santos et al. - in prep.),
(6) \citep{ribas05},
(7) \citep{mamajek08},
(8) \citep{stanfordmoore20},
(9) \citep{ayres21},
(10) \citep{telleschi05},
(11) \citep{crandall20},
(12a,b) \citep{ayres22,penz08a},
(13) \citep{brown_e22},
(14) \citep{gondoin20},
(15) \citep{rochapinto04},
(16) \citep{nissen15},
(17) \citep{baker05}.
  Uncertainties on the quoted L(EUV)/L$_{bol}$ are dominated by the RMS scatter on the \ion{N}{5}-to-EUV scaling relations (factor of $\approx$~2; \citealt{france18}) and modeled uncertainty on DEM calculations ($\approx$~30~--~50~\%; \citealt{duvvuri21}); uncertainties on the direct \ion{N}{5} fluxes are 5~--~10\% and do not contribute significantly to the total L(EUV)/L$_{bol}$ uncertainty.}
\end{deluxetable*}
\end{longrotatetable}

\subsection{EUV Scaling Relations}

In reaction to the dearth of observed EUV spectra of cool stars, numerous groups have developed scaling relations based on easier-to-observe proxies.  Scaling relations based on higher formation temperature (e.g., coronal) X-ray observations~\citep{sanzforcada11}, lower temperature, chromospheric tracers such as Ly$\alpha$~\citep{linsky14} and \ion{Ca}{2} H\&K~\citep{sreejith20}, and solar EUV/X-ray and magnetic flux relationships~\citep{chadney15,king18,namekata23} have been widely used in the literature (e.g., \citealt{youngblood16, airapetian17,oklopcic19}).  For the present work, we adopt the scaling relation presented by~\citet{france18}, who correlate direct $EUVE$ observations of 11 cool stars in the 90~--~360~\AA\ range with  $HST$ observations of the \ion{N}{5} doublet ($\lambda$$\lambda$1238, 1243~\AA; Figure 1).  While the majority of the EUV emission from cool stars is formed in the corona, \ion{N}{5} has the highest formation temperature ($\approx$ 1~--~2~$\times$~10$^{5}$ K) of commonly observed FUV emission lines, encompassing the transition region and the tails of the chromospheric and coronal plasma distributions.  The RMS scatter in the empirical \ion{N}{5}-to-EUV relationship is approximately a factor of 1.7 across F~--~M stars with varying activity levels~\citep{france18}, and provides an observationally-motivated proxy for the EUV luminosity of cool stars.   


Since the \ion{N}{5} scaling relations are only valid for the 90~--~360~\AA\ region where $EUVE$ spectroscopy exists, we use this wavelength region to represent the total EUV flux for all stars used in this work.  The 90~--~360~\AA\ range makes up $\gtrsim$ 50\% of the EUV flux from Sun-like stars and dominates the photoelectric heating of Earth-like atmospheres~\citep{youngblood25}. This is a similar constraint as chosen in~\citet{ribas05} and the range over which DEM-based EUV spectra are evaluated in the results below.  In Section 3.3, we present `correction factors' to compute the total 90~--~911~\AA\ EUV flux for young and old solar analogs.

\subsection{New and Archival Far-Ultraviolet Spectra with $HST$}
\label{sec:obs}

The sample of solar analogs was defined by stars having FUV spectra with the Cosmic Origins Spectrograph (COS) or Space Telescope Imaging Spectrograph (STIS) in the archive; we therefore present \ion{N}{5}-based scaling relation EUV luminosities for all sources.  To complement the \ion{N}{5}-based EUV flux estimates, we derive the EUV luminosity and spectral energy distributions for a subset of stars with high-quality X-ray and FUV observations in the literature using a differential emission measure (DEM) technique (see, e.g., \citealt{drake20,duvvuri21}, and see Section 2.4 for a description of the DEM method).  Not all sources have high-quality X-ray observations available in the literature, particularly for the older portion of the sample, therefore we include sources with DEM-based EUV fluxes as additional ``measurements" of the given source when available; these points also provide useful information about the spread of the recovered EUV fluxes with the different methods. 
When available, we also present the direct observations from $EUVE$ (Section 3.4; Table 3), corrected for interstellar attenuation. The $EUVE$ sources are similar to the original \citet{ribas05} list (as there has not been another EUV observatory since $EUVE$), with the difference that we separate $\pi$$^{1}$ Uma and $\chi$$^{1}$ Ori, and include $\alpha$ Cen A~\citep{france18}.   

We also leverage the young exoplanet discoveries from $TESS$ and recent young Sun programs to bolster the early evolutionary coverage: HIP 67522 was observed as part of the Mega-MEATS program (HST GO16701, PI: A. Youngblood, see \citealt{chia24} and Froning et al. - submitted), DS Tuc A was observed as part of $HST$ GO17305 (PI: K. France, see dos Santos et al. - submitted), and V1298 Tau was observed as part of $HST$ GO16163 (PI: P. Cauley, see \citealt{duvvuri23}).  In addition, new deep observations of EK Dra (GO17595, PI: K. France) and $\kappa$$^{1}$ Cet (GO15825, PI: D. Soderblom and analyzed as part of the Mega-MEATS program) are presented here for the first time.  

For the very youngest stars ($\leq$ 10 Myr), we analyzed archival observations from the $HST$ GO15310 (PI: C. Johns-Krull) Upper Sco survey.  The majority of the older stars in the sample come from the EclipSS Survey (GO 15300, PI: T. Ayres~\citealt{ayres21}).  All spectral data were acquired with the COS G130M,  STIS E140M, and STIS G140L modes using standard pipeline reduction processes.  Figure 1 shows example fractional \ion{N}{5} spectra of young (DS Tuc A), intermediate ($\kappa$$^{1}$ Cet), and old ($\alpha$ Cen A) stars.   The \ion{N}{5} spectra of all targets are very well detected, with \ion{N}{5} fluxes of the lowest signal-to-noise targets (those made with the lower-sensitivity echelle mode on $HST$-STIS) measured with $<$~6~\% uncertainty.  Therefore, for most stars, the absolute flux calibration of the $HST$ spectrographs ($\approx$~5\%) is the limiting factor on the \ion{N}{5} input data.  To conservatively account for the combined photometric, fit, and calibration uncertainties, we assign a 10\% total uncertainty to the $F$(NV) values in Table 2.

\begin{figure}[t]
   \centering
   \includegraphics[scale=.35,clip,trim=0mm 0mm 0mm 0mm,angle=0]{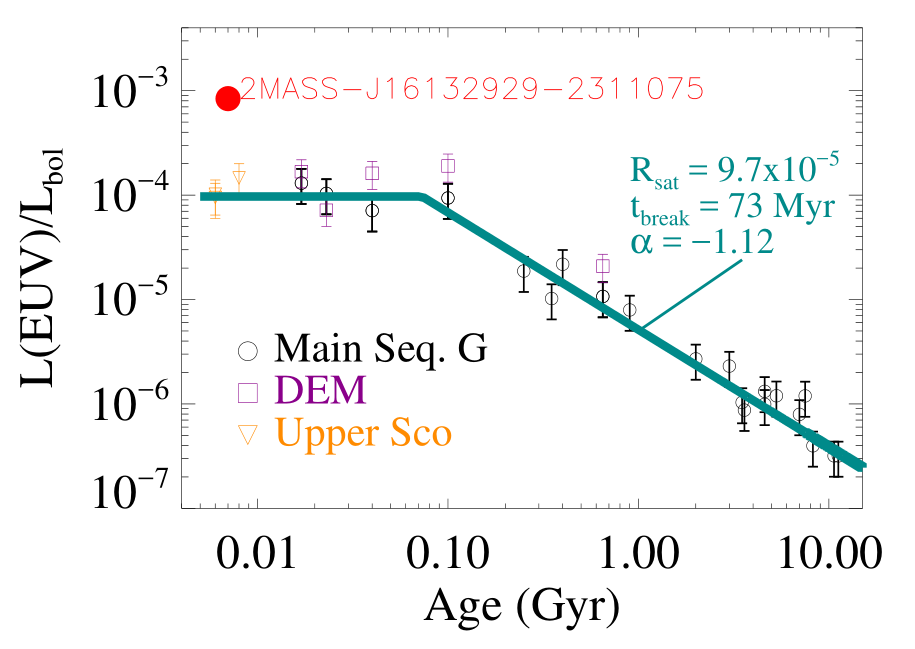}
   \figcaption{The fractional (90~--~360~\AA) EUV activity level of solar analogs as a function of time.  The data are comprised of pre-main sequence solar analogs located in the Upper Scorpius star-forming region (orange triangles), the young Suns with available DEM-based synthetic spectra in the literature (dark magenta open squares), and EUV luminosities derived from the \ion{N}{5}-based scaling relations developed by (\citealt{france18}; open black circles).  Uncertainties are defined as the RMS scatter on the \ion{N}{5}-to-EUV scaling relations (factor of $\approx$~1.7) and modeled uncertainty on DEM calculations ($\approx$~30~--~50~\%; \citealt{duvvuri21}).
   The best fit piecewise evolution model (Equation 3) is overplotted as the solid, dark cyan line, with best fitting parameters shown in the dark cyan legend text. The 7 Myr Upper Sco star Upper Sco 2MASS J16132929-2311075 (filled red circle) is an outlier; we discuss this star in the Appendix.}
\end{figure}

\subsection{Emission Line Data Analysis and EUV Estimates}

The \ion{N}{5} emission lines in our sample were directly measured from $HST$-COS and STIS spectra or taken from~\citet{ayres21}.  Table 1 references the emission line fluxes that were taken from previous observations.  For the new emission line analyses, we used an interactive Gaussian fitting routine where one or two Gaussian components (depending on the stellar activity level, see, e.g., \citealt{wood97}) were fitted to each member of the \ion{N}{5} doublet ($\lambda_{rest}$ = 1238.804 and 1242.795~\AA, respectively), and the total flux is the addition of the two lines.  The spectral line-fitting routine is based on a Levenberg-Marquardt minimization routine where the model Gaussian line shape is convolved with the instrumental line-spread-function for comparison with the data~\citep{france12}.  For the most active stars, we screened the $HST$ spectra for flares, finding several bright emission-line flares in the spectra of HIP 67522 (Froning et al.~--~submitted), DS Tuc A (Mento et al.~--~in prep.), and EK Dra (Namekata et al.~--~in prep.), however, the flares are of sufficiently short duration and the basal fluxes of the young solar-type star are sufficiently high, that the inclusion or exclusion of flares does not significantly change ($\lesssim$~10~\%) the total \ion{N}{5} fluxes presented here.

The \ion{N}{5} fluxes were then divided by the bolometric fluxes for each target assuming 
\begin{equation}
F_{bolom}= \sigma T_{eff}^4 \left( \frac{R_{\ast}}{d} \right)^2
\end{equation}
where the $T_{eff}$ is the stellar effective temperature, $R_{\ast}$ is stellar radius, and $d$ is the distance to the star. The bolometric flux for each source is presented in Table 1.  The 90~--~360~\AA\ fractional luminosity for the \ion{N}{5} sample is then 
\begin{equation}
log_{10} \left(L(EUV) / L_{bol} \right) = m \times log_{10} \left(F(NV) / F_{bol} \right) + b
\end{equation}
for the 90~--~360~\AA\ EUV, taking [$m$,$b$] = [1.00, 1.91]~\citep{france18}.
For the sources with DEM spectral estimates, we take the synthetic EUV spectrum and integrate the 90~--~360~\AA\ directly to compute L(EUV)/L$_{bol}$.

 We complement the \ion{N}{5}-based EUV estimates with new and archival synthetic spectra derived from DEM analyses. 
Stellar FUV and X-ray emission span a broad range of formation temperatures, providing good constraints on emission formed between 10$^{4}$~--~10$^{5.2}$ K and 10$^{6.2}$~--~10$^{7}$ K, respectively, and these observations constrain the temperature-dependent contribution function and differential emission measure function. 
To derive the emission from the intermediate temperature plasma ($10^{5.2-6.2}\,K)$, the differential emission measure is approximated with a non-parametric function~\citep{kashyap98,plowman13} or polynomial~\citep{duvvuri21} and evaluated at the respective temperatures; this intermediate and high-temperature temperature plasma contributes  a large fraction of the 100 – 911 \AA\ EUV luminosity ($>$~80\% at $\lambda$~$\leq$~400~\AA\ and 20~--~50\% from 500~--800~\AA, e.g., \citealt{tilipman21}).   Figure 2 shows a sample of these for young (DS Tuc A) and intermediate ($\kappa$$^{1}$ Cet) sources, and includes the present day Sun at an intermediate activity level (F$_{10.7}$~$\approx$ 140 sfu; 1 sfu = 10$^{-19}$ erg s$^{-1}$ cm$^{-2}$ Hz$^{-1}$) for comparison.  These spectra are shifted to the flux density at 1 AU to illustrate the dramatic change in EUV flux in the habitable zone over the lifetime of a solar-type star.


\section{Discussion: The Evolution of the Solar EUV Luminosity}

\begin{figure}[t]
   \centering
   \includegraphics[scale=.37,clip,trim=0mm 0mm 0mm 0mm,angle=0]{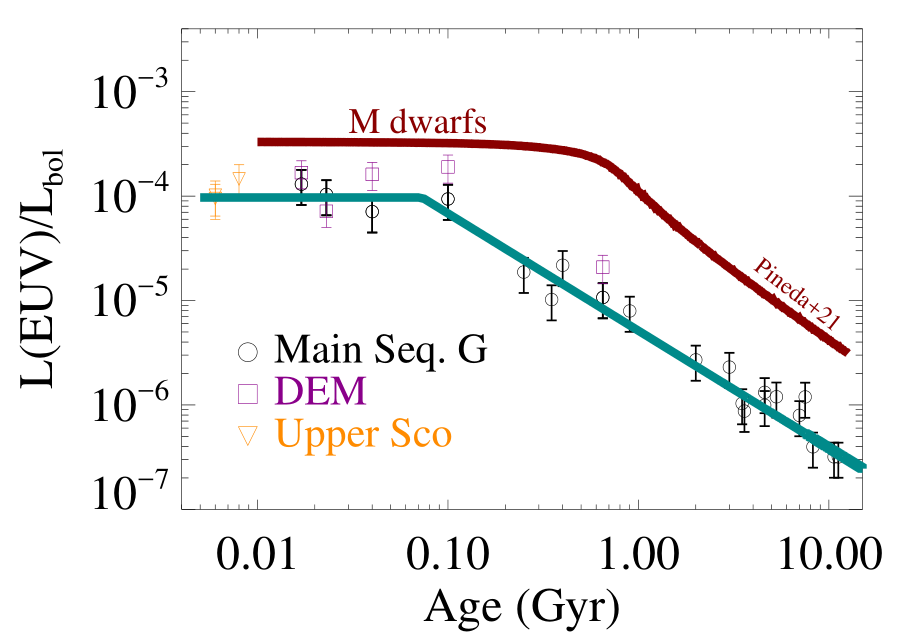}
   \figcaption{Same as Figure 3 (fractional (90~--~360~\AA) EUV activity level of solar analogs as a function of time), but including the comparable early-M dwarf EUV activity evolution from~\citet{pineda21}.  The integrated area in the M dwarf EUV history is a factor of $\approx$~10 higher than the solar case. }
\end{figure}

\subsection{EUV Activity Fraction}

The resultant L(EUV)/L$_{bol}$ observations are shown as a function of age in Figure 3, with the different subsamples coded by color.  As expected, we observe that the EUV activity follows a saturated early phase transitioning to an EUV activity decline after a characteristic breakpoint time, $t_{break}$.     


We fitted the EUV activity-age diagram with a power-law of the form: 
\begin{equation}
L(EUV)/L_{bol} = \begin{cases} R_{sat}, &  t < t_{break}  \\  R_{sat} \times (t/t_{break})^{\alpha},  &  t~\geq~t_{break}  \end{cases} 
\end{equation}
where $R_{sat}$ is the saturated activity level, $t$ is the age in Gyr, t$_{break}$ is the breakpoint time (also in Gyr) where the EUV activity declines, and $\alpha$ is the power-law slope. 
The best fit parameters for the three model parameters are $R_{sat}$ = 9.7 ($\pm$ 1.6)~$\times$~10$^{-5}$, $t_{break}$ = 73 $\pm$ 16 Myr, and $\alpha$ = $-$1.12~$\pm$~0.06.  We reiterate here that the above results assume that the EUV activity uncertainty is dominated by the EUV calculations; we do not consider age or stellar property uncertainties since a uniform set of measurements is not available for solar analogs spanning this age range with UV data.  Building a comprehensive database of solar analogs systematically studied in the optical for rotation periods and metallicities, and the UV/X-ray for the transition region and coronal contributions (or better, directly in the EUV, Section 4), may offer significant improvement upon the present study.     

We note that while $t_{break}$ is technically determined to $\sim$~20\% accuracy, the lack of stars in the break region from 50~--~250 Myr makes the retrieved value sensitive to the exact ages of individual stars (specifically EK Dra, $\pi^{1}$ UMa, and $\chi^{1}$ Ori in this study). Consequently it is difficult to clearly identify the EUV activity breakpoint age.  In addition, the early activity evolution of solar-type stars is expected to have a dependence on the initial rotation~\citep{johnstone21}, therefore additional \ion{N}{5} observations of young Suns in this age range would be very valuable.  

In Figure 4, we compare our EUV activity history with the equivalent, \ion{N}{5}-derived curve for early M dwarfs from~\citet{pineda21}.  Several differences are immediately clear. First, the breakpoint time for the M dwarf activity is significantly later than for solar type stars ($\sim$~1 Gyr vs 75 Myr), as has been noted by numerous previous papers (e.g., \citealt{loyd21,engle24,pass25}). Second, while there is scatter around the average value for both G- and M-type stars, the M dwarf EUV saturation level is approximately 0.5 dex higher.  
We find that early M dwarfs deposit a factor of $\approx$11 times more EUV energy at a given orbital distance over the full 10 Myr~--~10 Gyr range.  This value is consistent with previous M dwarf studies that include the effects of bolometric luminosity evolution; \citet{pineda21} find a factor $\sim$~10~---~20 more EUV energy delivered at a given orbital distance for early-to-mid M dwarfs and \citet{ribas16}, considering the full XUV irradiance, find that Proxima Cen deposits 7~--~16 times more high-energy radiation over its $\sim$5 Gyr lifetime.   

\begin{figure}[t]
   \centering
   \includegraphics[scale=.3,clip,trim=0mm 0mm 0mm 0mm,angle=0]{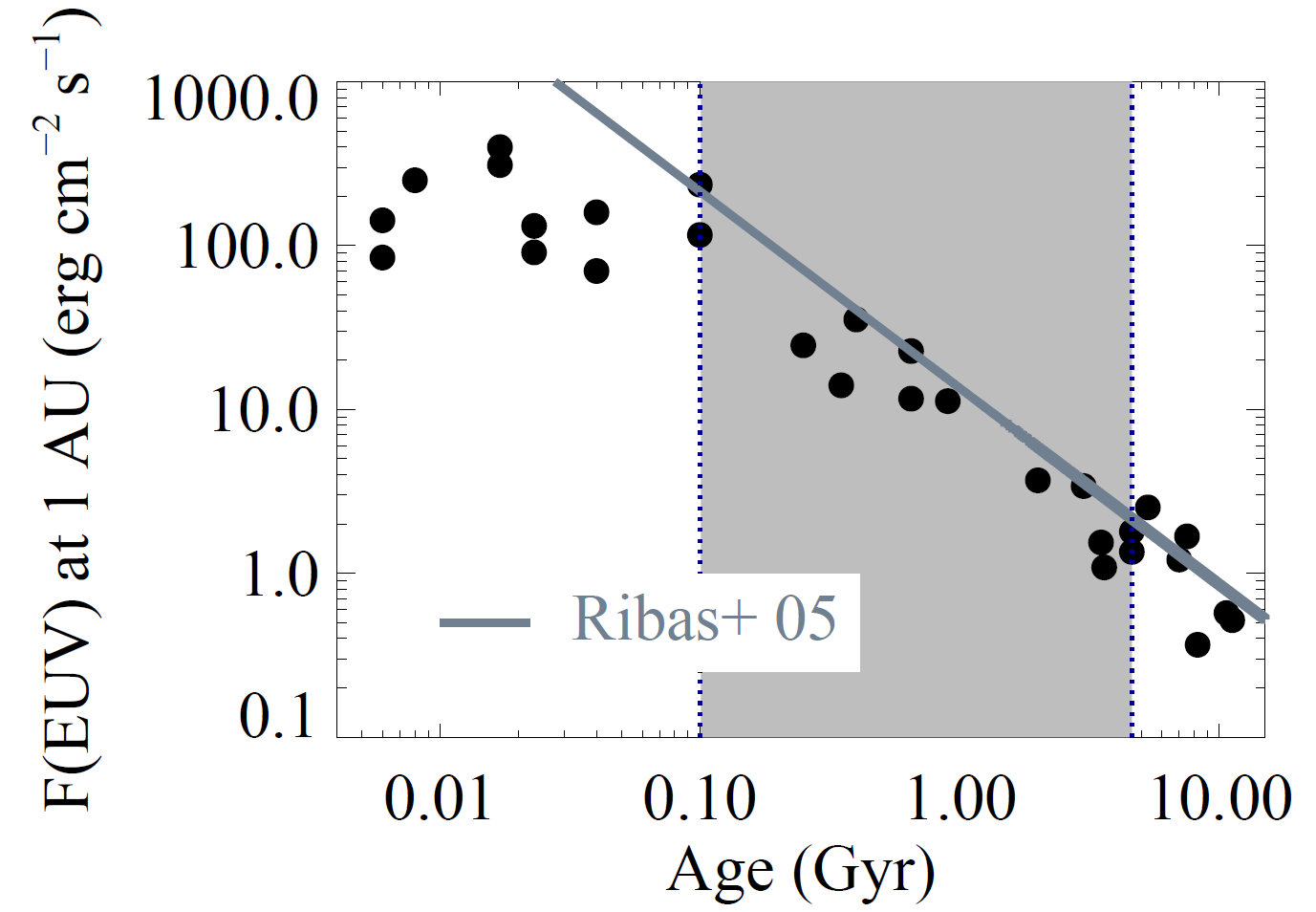}
   \includegraphics[scale=.41,clip,trim=0mm 0mm 0mm 0mm,angle=0]{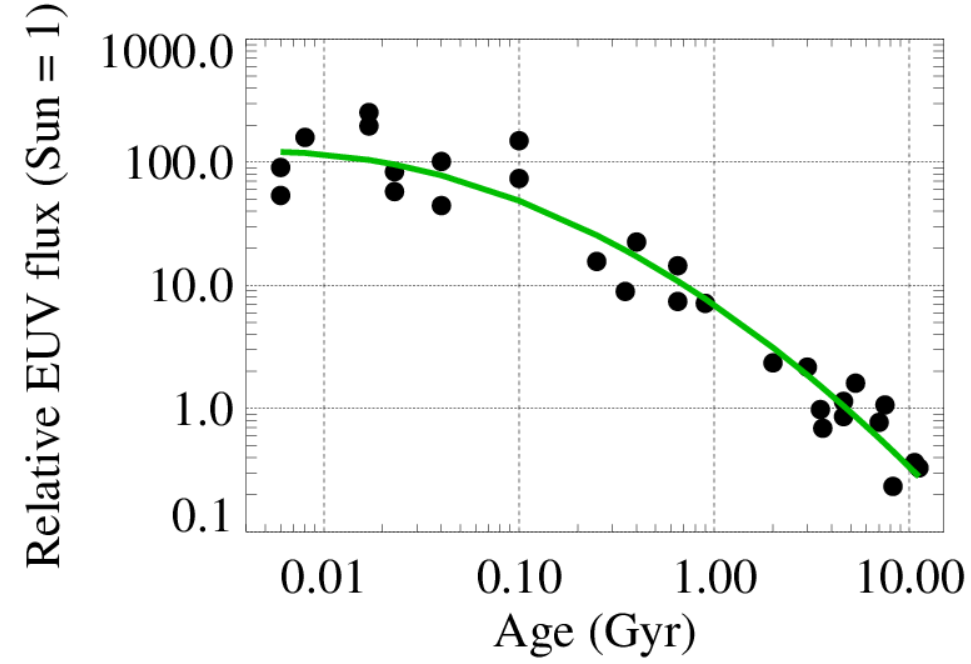 }
      \figcaption{The $top$ plot shows the 90~--~360~\AA\ EUV flux at 1 AU from the central star.  The dark gray curve shows the EUV flux evolution fit from~\citet{ribas05}.  The 100 Myr~--~4.6~Gyr region over which Ribas et al. had empirical EUV data are encompassed by the shaded gray area.  At bottom, we show the same plot, normalized to the present-day solar value for intermediate activity conditions (F$_{\odot}$(EUV) = 1.57 erg cm$^{-2}$ s$^{-1}$).  The green curve shows the best-fit to the normalized EUV flux, demonstrating that the relative EUV flux experienced within the habitable zone varies from approximately 100~$\times$ to 0.3~$\times$ over the Sun's 10 Myr~--~10 Gyr lifetime.     }
\end{figure}

\subsection{EUV Flux Evolution at 1 AU}

Figure 5 presents the absolute and relative stellar EUV flux incident at 1 AU from the star.  Figure 5 ($top$) shows the data points from this study compared to the best fit 100~--~360~\AA\ power-law curve fit to the 4 stars in the~\citet{ribas05} sample.  We find that our points are bounded by the  Ribas et al. curve over the 0.1~--~4.6 Gyr range considered in that study.  Our points are somewhat lower than the Ribas et al. curve on average ($\approx$~30~--~35\%), with the points falling off significantly at ages below 0.1 Gyr where the saturated EUV regime, not covered in the Ribas et al. work, flattens the activity curve.  

We can divide the 1 AU flux points by the average of the present day solar 90~--~360~\AA\ flux for the MIN and MAX solar cases presented in Table 1, F$_{\odot}$(EUV) = 1.57 erg cm$^{-2}$ s$^{-1}$, to present a relative EUV scale factor for the Sun as a function of time (Figure 5, $bottom$).  In these units, the present-day average solar EUV flux has a value of unity and we parameterize this function as a quadratic: 
\begin{equation}
  \mathcal{F}_{rel}(EUV)  = a_{2} \times \tau^2 + a_{1} \times \tau + a_{0} 
\end{equation}
where $\tau$~=~log$_{10}$($t$) and 
$\mathcal{F}$$_{rel}$(EUV) = log$_{10}$(F$_{*}$(EUV)/F$_{\odot}$(EUV)).  The solar EUV flux received was at a maximum of ~$\sim$~100$\times$ its present value at 10 Myr (soon after dissipation of the Sun's protoplanetary disk) and will decrease to $\sim$~30\% of its present value by $\sim$~10 Gyr.   The coefficients from the best fit line in Equation 4 are: [$a_{0}$ = 0.84~$\pm$~0.07, $a_{1}$ = $-$1.08~$\pm$~0.07, $a_{2}$ = $-$0.24~$\pm$~0.05].  We find that 50\% of the lifetime integrated EUV flux is delivered to orbiting planets by $\approx$~550 Myr, consistent with the results of~\citet{king21} who found that the large majority of the escape-driving EUV flux from solar-type stars is delivered after the star has left the saturated activity regime.  In contrast, M dwarfs are believed to deliver the majority of their EUV or XUV energy by $\sim$~1 Gyr, while the stars are in the saturated regime~\citep{pineda21,pass25}.

\begin{figure}[htbp]
   \centering
   \includegraphics[scale=.36,clip,trim=0mm 0mm 0mm 0mm,angle=0]{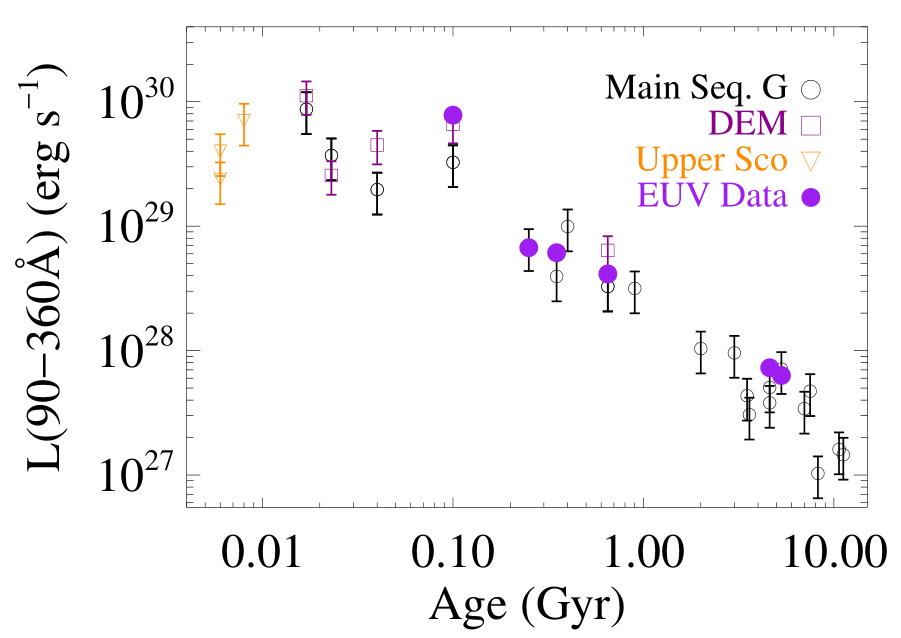}
      \includegraphics[scale=.36,clip,trim=0mm 0mm 0mm 0mm,angle=0]{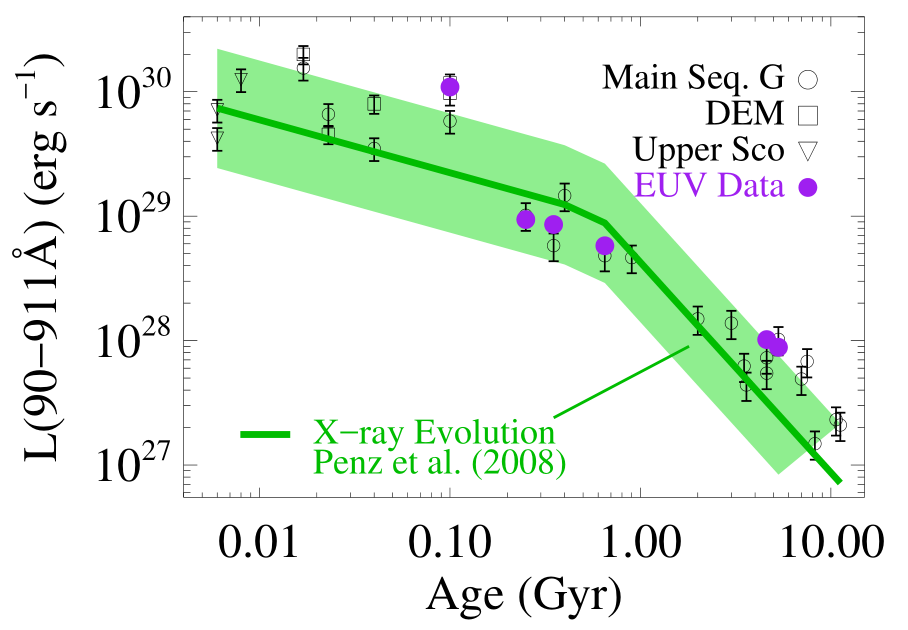}
      \figcaption{ ($top$) The 90~--~360~\AA\ luminosities of each of the stars in our survey as a function of stellar age.  Symbols are the same as in Figure 3, with the addition of direct EUV spectral observations from $EUVE$ (integrated over the 90-360~\AA\ bandpass following a correction for interstellar hydrogen and helium absorption; purple filled circles).   The $bottom$ plot shows the complete 90~--~911~\AA\ EUV luminosities (described in Section 3.3).  The G-star X-ray history from \citet{penz08a} employed in the Cosmic Shoreline formulation~\citep{lammer09, zahnle17} is shown overplotted as the solid green line (approximate uncertainty range shown as the shaded green region).  We see that the EUV and X-ray luminosities are comparable for stars at ages~$\lesssim$~1 Gyr, after which the X-rays decline and the EUV luminosity dominates (see also~\citealt{king21}).   }
\end{figure}

\subsection{Correction Factors for the Full 90~--~911\AA\ EUV Bandpass}

We present `correction factors' to convert the 90~--~360~\AA\ EUV range considered in this work to the full 90~--~911~\AA\ band (noting that we are deviating slightly from the typical EUV wavelength definition, 100~--~911~\AA, to accommodate the \ion{N}{5}-based scaling relation).  We compute the full EUV correction factors for representative young ($<$~100 Myr), intermediate (100 Myr~--~1 Gyr), and older ($>$~1 Gyr) Suns by measuring the 90~--~360~\AA\ and 360~--~911~\AA\ from the six spectra with full EUV wavelength coverage (see, e.g., Figure 2).   These correction factors, based on DEM calculations for the young and intermediate categories, and based on the Sun for the older category, are given in Table 2.  We use the average correction factor of HIP 67522, V1298 Tau, and DS Tuc A, for the young category and the average correction factor of EK Dra and  $\kappa^{1}$ Cet for the intermediate category.   While the small numbers do not permit us to identify clear trends, the correction factors suggest a higher correction factor for the youngest stars, with the average correction factor settling to $\approx$~1.4~--~1.5 after 100 Myr. An outlier to this trend is DS Tuc A, which shows the lowest correction factor (1.34) in the sample.   

The correction factors vary less than 35\% from $\approx$~17 Myr to the modern day Sun (4.6 Gyr). Even though the spectra are qualitatively different (Figure 2), on a band-integrated level, they do not change dramatically.  The emission line-to-continuum ratio changes with coronal temperature owing to the formation temperature differences between the hot coronal emission lines and the atomic continua.  More active stars (hotter coronae) increase the emission line fluxes at longer EUV wavelengths, while less active stars have a larger relative contributions over 360~--~911~\AA\ from the \ion{H}{1} and \ion{He}{1} continua.  We suggest that the balance in the relative contributions of the different spectral components in the young and old stars conspire to prevent strong evolutionary changes in the (90~--~360~\AA)/(360~--~911~\AA) flux ratio.

\begin{table*}[t]
\centering
\caption{90~--~360~\AA\ to 90~--~911~\AA\ EUV Correction Factors$^{a}$ } 
\begin{tabular}{ccc}
\hline \hline
Star &  F(90~--~911~\AA)  &    90~--~911\AA\ Correction Factor \\ 
     &  (erg cm$^{-2}$ s$^{-1}$)  &    \\   
\hline
HIP 67522   &    1.18E-12    &     1.96  \\
V1298 Tau   &    3.71E-13    &     2.03  \\
DS Tuc A   &    2.58E-12    &     1.34  \\
EK Dra   &    6.64E-12    &     1.42  \\
$\kappa^{1}$ Cet   &    9.44E-12    &     1.52  \\
Intermediate Activity Sun$^{b}$   &    3.73E+00    &     1.44  \\
\hline
   &    &    \\ 
\hline
Age  Range &  90~--~911\AA\ Correction Factor  &  Stellar Proxy  \\ 
\hline
Young ($<$~100 Myr)	& 1.78 &  $<$ HIP 67522, V1298 Tau, DS Tuc A $>$	  \\
Intermediate	(100 Myr~--~1 Gyr) & 1.47 &  $<$ EK Dra, $\kappa^{1}$ Cet $>$  \\
Older ($>$~1 Gyr)  & 1.44  &   Intermediate Activity Sun$^{b}$   \\
\hline
$^{a}$ F(90~--~911~\AA) = F(90~--~360~\AA)~$\times$~(Correction Factor)\\
$^{b}$ F$_{10.7}$ = 140
\end{tabular}
\vspace*{-0.0in}
\end{table*}


\subsection{EUV vs X-ray Luminosity Evolution of Solar-type Stars}

Finally, we consider the absolute EUV luminosity evolution and compare it to the average X-ray luminosity for G-type stars.  The 90~--~360~\AA\ EUV luminosity is shown in Figure 6 ($top$) with the youngest stars located between 10$^{29}$ and 10$^{30}$ erg s$^{-1}$, a similar break near 100 Myr, and a decline to L(90~--~360~\AA)~$\sim$~10$^{27}$ erg s$^{-1}$ by 10 Gyr.  While the luminosity for the oldest solar-type stars is more than order of magnitude higher than for a comparably old M dwarf~\citep{france20}, when normalized to the orbital distance, this flux is significantly lower than old red dwarfs.   

 We also compare the predicted 90~--~360~\AA\ EUV luminosity with direct observations from $EUVE$ (EK Dra, $\pi^{1}$ Uma, $\chi^{1}$ Ori, $\kappa^{1}$ Cet, and $\alpha$ Cen A; Table 3). $EUVE$ fluxes of  $\chi^{1}$ Ori, $\kappa^{1}$ Cet, and $\alpha$ Cen A were taken directly from \citet{france18}. We performed the same analysis for $\pi^{1}$ Uma, using the \ion{H}{1} Ly$\alpha$ observations from Wood et al. (2014; log$_{10}$(N(HI)) = 18.12)\nocite{wood14} to constrain the interstellar column density.  EK Dra is marginally detected by $EUVE$ outside of a small number of coronal lines and therefore the integrated 90~--~360~\AA\ flux is sensitive to the exact background subtraction applied in the reduction.  For EK Dra, we instead adopted the EUV luminosity from~\citet{johnstone21}, corrected for the updated interstellar column density from Youngblood et al. (2025; log$_{10}$(N(HI)) = 18.01)\nocite{youngblood25}.  We find that the observed $EUVE$ luminosities are mostly consistent within 1- to 2-$\sigma$ of the \ion{N}{5}-based scaling and DEM-based EUV calculations.

\begin{table*}[t]
\centering
\caption{90~--~360~\AA\ Fluxes from $EUVE$ } 
\begin{tabular}{lccc}
\hline \hline
Star &  log$_{10}$(N(HI)) &  F(90~--~360~\AA)  &    Reference \\ 
         &               &  (erg cm$^{-2}$ s$^{-1}$)  &    \\   
\hline
EK Dra   &   18.01     &   5.5E-12    &         \citep{johnstone21}$^{a}$ \\
$\pi^{1}$ Uma   &   18.12     & 2.7E-12    &       this work   \\
$\chi^{1}$ Ori  &   17.93     & 6.8E-12    &          \citep{france18} \\
$\kappa^{1}$ Cet   &   17.89     & 4.0E-12    &       \citep{france18}   \\
$\alpha$ Cen A &  17.61     &  2.9E-11    &       \citep{france18}  \\
\hline
$^{a}$ Updated column density to \citet{youngblood25} value.
\end{tabular}
\vspace*{-0.0in}
\end{table*}


Figure 6 ($bottom$) displays the full 90~--~911~\AA\ EUV luminosity, after correcting for the longer-wavelength EUV (see previous subsection).  We overplot the G-dwarf X-ray evolution curve from~\citet{penz08a}, which is the ROSAT X-ray history assumed in the current formulation of the Cosmic Shoreline~\citep{lammer09,zahnle17}.   While noting the significant scatter associated with the limited sample size (and possibly reflecting differences in the initial stellar rotation rates; \citealt{johnstone21}), we find that the L(EUV) and L(X-ray) curves are in general agreement for ages $<$~1 Gyr, suggesting approximate energy equipartition between the two bands  (with a suggestion that the EUV luminosity is higher than the X-ray for ages $\leq$~100 Myr).  At ages older than $\sim$~1 Gyr, we see the EUV luminosity exceed the X-ray luminosity by factors of a few, approximately consistent with the predictions of ~\citet{king21}.

\section{Conclusions}

In the preceding sections, we have employed \ion{N}{5}-EUV scaling relations and DEM calculations to present a new evolutionary history of the EUV activity on solar-like stars.  We take advantage of the large investment in FUV spectroscopy of cool stars with $HST$ in the 20 years following the pioneering Sun in Time study~\citep{ribas05}, and specifically the sensitivity increase provided by the $HST$-COS instrument to provide spectra of not only active but distant and older (less active) solar analogs.  Our sample covers archival observations of Weak Line T Tauri stars in Upper Sco, new observations of Young Suns, and archival spectra from the EclipSS survey of intermediate-to-old solar analogs~\citep{ayres21}.     

We can summarize our findings in the following points: 
\begin{enumerate}
    \item The EUV activity evolution for solar-type stars follows a two-component behavior: a saturated L(EUV)/L$_{bol}$ plateau (at a level of about 10$^{-4}$) followed by a power law decay (characterized by a power-law slope, $\alpha$ of approximately $-$1.1). This behavior, expected based on age-activity relations observed at other wavelengths, is seen for the first time in the solar EUV history.  We find a break from EUV activity saturation around 50~--~100 Myr, but note that sparse age coverage around the break prevents a firm measurement of the saturation duration.  We suggest that additional $HST$-COS observations of solar analogs, particularly those targeted towards gaps in the stellar age distribution presented here, would be very valuable. Similarly, ground- or space-based rotation period studies for older solar type stars would provide another valuable direction to expand this age-activity study.   
    
    \item We compare the solar EUV activity with similar estimates from early/mid-M dwarfs, finding that the M dwarf saturation level is several times higher and lasts $\gtrsim$10~--~20 times longer.

    \item The EUV flux in the habitable zone is computed as a function of stellar age.  We find that the habitable zone EUV flux, relative to the present day value on Earth, changes from 100~$\times$ at 10 Myr to 0.3~$\times$ at 10 Gyr.  A functional form of this curve is presented in Equation 4.  

    \item The total 90~--~911~\AA\ EUV luminosity history is presented and compared with the soft X-ray luminosity of G stars.  We find that the EUV and soft X-ray evolution are consistent within the uncertainties from 10 Myr to 1 Gyr, with the EUV luminosity dominating in older systems.   
    
\end{enumerate}

An important caveat to this work is that, except for the small number of purple points in Figure 6, there is {\it no direct EUV data} in this study of EUV evolution.  Unlike the well-characterized X-ray behavior of F, G, K, and M stars, our archive of high-quality EUV spectra of low-mass stars is limited to the $\sim$10 nearby F, G, and active M stars observed by $EUVE$~\citep{drake95,drake97,ribas05,craig97}.  Consequently, the present day EUV output of a given star, its evolution, flare behavior, and association with stellar CMEs, is only as well defined as our spectral approximations and sparse FUV or X-ray samples, such as presented in this work.  The proposed {\it Extreme-ultraviolet Spectral Characterization for Atmospheric Physics and Evolution} mission concept ($ESCAPE$; \citealt{france22}) is designed with $\sim$100$\times$ the spectral sensitivity of $EUVE$ to map these quantities as a function of stellar mass.  High-sensitivity EUV spectroscopy of a broad sample ($>$ 200) of nearby stars would enable us to understand the EUV history of the Sun, how it compares to stars of other masses, and how this history impacts the temperate, terrestrial planets studied by $JWST$, $HWO$, and future missions.  

\acknowledgments
  KF thanks members of the Rocky Worlds HST+JWST program SAC and CIT for engaging discussion during the development of this paper. This work was part of $HST$ GO programs 16701, 17305, and 17595, and supported by the associated grants to the University of Colorado at Boulder. V.S.A. was supported by the GSFC Sellers Exoplanet Environments Collaboration (SEEC), which is funded by the NASA Planetary Science Division’s Internal Scientist Funding Model (ISFM), NASA’s Astrophysics Theory Program grant 80NSSC24K0776NNH21ZDA001N-XRP, F.3 Exoplanets Research Program grants, NICER Cycle 2 project funds and NICER DDT program.  Archival data used in this paper were obtained from the Mikulski Archive for Space Telescopes (MAST) at the Space Telescope Science Institute. 
  Some of the data presented in this article were obtained from the Mikulski Archive for Space Telescopes (MAST) at the Space Telescope Science Institute. The specific observations analyzed can be accessed via \dataset[doi:10.17909/g701-t514]{http://dx.doi.org/10.17909/g701-t514}.

\bibliography{Main_Astroph_EUVevol_v1}{}
\bibliographystyle{aasjournal}

\appendix 
\section{Upper Sco 2MASS J16132929-2311075: Declining Phase of a Pre-Main Sequence Superflare?
}

When creating the new EUV activity time evolution curves, we found that 2MASS J16132929-2311075 was a strong outlier to the otherwise relatively well-behaved curve, sitting at L(EUV)/L$_{bol}$ ~$\approx$~10$^{-3}$, an order of magnitude higher than other stars in the Upper Sco region (Figure 3). However, unlike the older solar analogs in our sample, corrections for dust reddening are important for the Upper Sco stars.  Upon visual inspection, it was clear that the 2MASS J16132929-2311075 spectrum was qualitatively of much higher S/N than any of the other pre-solar Upper Sco stars in the GO15310 sample (Figure 7).  This indicates that the high EUV activity inferred for this star was not solely due to an overestimated extinction correction. We analyzed the COS spectrum in more depth to understand the inferred strong EUV emission from 2MASS J16132929-2311075.

 The strong \ion{N}{5} lines are reminiscent of the profiles seen in accreting young stars~\citep{france14}, and other Sco stars are known to maintain long-lived gas disks and mass accretion~(e.g., \citealt{castro16}), however, a careful examination shows no sign of the fluorescent H$_{2}$ signature that is the far-ultraviolet signpost for remnant gas disks~\citep{ingleby11b,france12,alcala19}.  
 Inspection of the remainder of the 2MASS J16132929-2311075 spectrum revealed strong \ion{Fe}{21} emission (Figure 7), which has previously been used as an indication of the presence of strong coronal flares in the FUV spectra of cool stars~\citep{froning19,france20}.  The stellar \ion{N}{5} and \ion{C}{2} were sufficiently bright to enable a reasonably high-quality lightcurve to be created from the $\approx$~2200 second observation (Figure 8).  The far-UV emission line lightcurves shows a shape consistent with an exponential decline in the 60-second cadence data, dropping by $\approx$~30\% over the course of the exposure.  This suggests that the stellar flux was declining from a previous flare event that peaked prior to the start of the COS observations.

\begin{figure*}[htbp]
   \centering
   \includegraphics[scale=.75,clip,trim=0mm 0mm 0mm 0mm,angle=0]{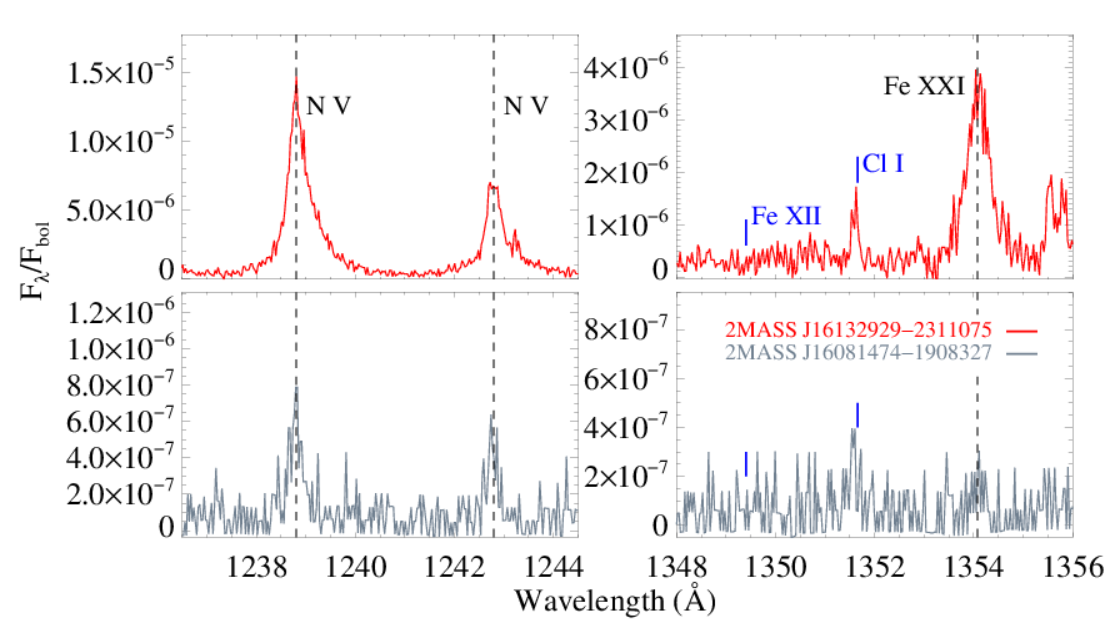}
   \figcaption{A comparison of two pre-main sequence stars in Upper Sco, Upper Sco 2MASS J16132929-2311075 (age~$\sim$~7 Myr, M$_{*}$~$\sim$~1.3 M$_{\odot}$; red histogram) and 2MASS J16081474-1908327 (age~$\sim$~6 Myr, M$_{*}$~$\sim$~1.3 M$_{\odot}$; gray histogram).  The `as observed' spectra are normalized by the bolometric fluxes of each star~\citep{pecaut16}; the exposure times are approximately equal for these two observation, but the differences in S/N are obvious. 2MASS J16132929-2311075 displays much larger relative hot gas fractions in \ion{N}{5} (left panels; T$_{form}$~$\approx$~1~--~2~$\times$~10$^{5}$ K) and \ion{Fe}{21}~(right panels; T$_{form}$~$\approx$~10$^{7}$ K).  We speculate that 2MASS J16132929-2311075 was observed in the declining phase of large stellar flare (Figure 8).}
\end{figure*}

 Given that we only have a brief snapshot of the decay phase, and because it is tangential to the goals of this work, we do not attempt to derive the detailed properties of this flare.  However, we consider a simple analysis where we assume all of the flux in the observation is contributed by a flare (F$_{130}$ = 4.03~$\times$~10$^{-13}$ erg cm$^{-2}$ s$^{-1}$, over the 1080~--~1365~\AA\ bandpass) and multiply by the distance and exposure time (4$\pi$$d^{2}$ $\times$~$t_{exp}$; $d$~=~138.5 pc; $t_{exp}$ = 2200 seconds), we find an FUV energy of $\approx$~2~$\times$~10$^{33}$ erg.  Considering that this is a very small slice of the stellar spectrum, and that we are excluding Ly$\alpha$ in this region (which will likely contribute an additional factor of 5~--~10 to the total energy; \citealt{france13}), the total bolometric energy of this event is likely larger than 10$^{34}$ ergs.  We consider it plausible that we are indeed observing the tail of an superflare~\citep{doyle20}.   We speculate that the combination of transition region and coronal activity an order of magnitude higher than other young stars, the steady flux decline seen in the lightcurves, and the total FUV energy $>$~10$^{33}$ erg strongly suggests that we observed the afterglow of a large pre-main sequence X-ray/UV superflare on 2MASS J16132929-2311075.

\begin{figure}[htbp]
   \centering
   \includegraphics[scale=.60,clip,trim=0mm 0mm 0mm 0mm,angle=0]{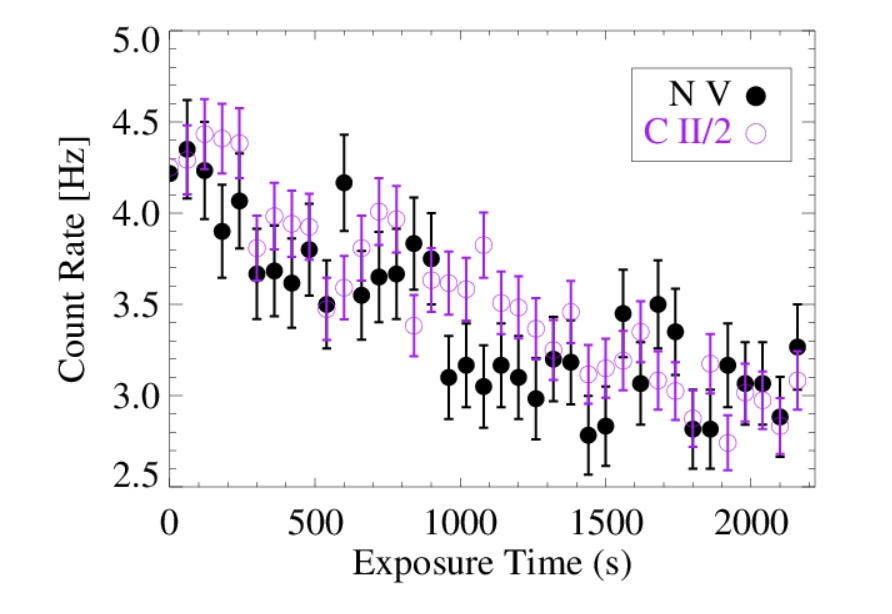}
   \figcaption{Emission line lightcurves of 2MASS J16132929-2311075 in the \ion{N}{5} and \ion{C}{2} lines.  The \ion{C}{2} count rates have been divided by two for display.  Both lines show a $\sim$ 30~--~40\% flux decrease across a single $HST$ orbit, suggesting that this star was observed in the declining phase of large stellar flare. }
\end{figure}

\end{document}